\documentclass[twocolumn,showpacs,prbprintnumbers,pre,fleqn,superscriptaddress]{revtex4}
\usepackage{epsfig}
\usepackage{graphicx}
\usepackage{amsfonts}
\usepackage[dvips]{color}

\newcommand{\ct}{\cite}
\newcommand{\bi}{\bibitem}
\newcommand{\be}{\begin{equation}}
\newcommand{\ee}{\end{equation}}
\newcommand{\ba}{\begin{eqnarray}}
\newcommand{\ea}{\end{eqnarray}}

\newcommand{\non}{\nonumber}
\newcommand{\bra}[1]{\langle #1|}
\newcommand{\ket}[1]{|#1\rangle}

\begin{document}
\title{ Quantum Discord in a spin-1/2 transverse XY Chain Following a Quench} 
\author{Tanay Nag}
\email{tanayn@iitk.ac.in}
\author{Ayoti Patra}
\email{ayoti@iitk.ac.in}
\author{Amit Dutta}
\email{dutta@iitk.ac.in}
\affiliation{Department of Physics, Indian Institute of Technology Kanpur,
Kanpur 208 016, India} 

\begin{center}
\begin{abstract}
We report a study on the zero-temperature quantum discord as a measure of two-spin correlation of a transverse $XY$ spin chain following a quench across a quantum critical point and
 investigate the behavior of mutual information,
classical correlations and hence of discord in the final state as a function of the rate of quenching. We show that though  discord vanishes in the limit of very slow as well as very fast
 quenching, it exhibits a peak for an intermediate value of the quenching rate. We show that though  discord and also the mutual information exhibit a similar behavior with respect to
the quenching rate to that of concurrence or negativity following an identical quenching, there are quantitative differences. Our studies indicate that like concurrence, discord also 
exhibits a power law scaling with the rate of quenching in the limit of slow quenching though it may not be expressible in a closed power law form. We also explore the behavior of  discord on quenching linearly 
across a quantum multicritical point (MCP) and observe a scaling similar to that of the defect density.
 
\end{abstract}
\end{center}

\maketitle

\section{Introduction}
 Quantum Phase Transitions(QPT) \ct{sachdev99, chakrabarti96,continentino,vojta03} driven by quantum fluctuations arising 
due to the change of a parameter in the Hamiltonian at absolute zero temperature have been studied extensively. A QPT
is characterised by a fundamental change in the symmetry of the ground state of a quantum many-body system and is also associated with diverging correlation length 
as well as a diverging relaxation time at the Quantum Critical Point(QCP). Over last few years, numerous efforts have been directed to
 understanding the connection between Quantum Information and QPTs \ct{osterloh02,osborne02,guo05,amico08}.
In recent years, QPTs have been
observed experimentally in a large number of systems, for example in optical lattices where a Mott insulator to superfluid transition is
observed \ct{makhlin00,greiner02,jaksch98}.

  The entanglement between two spins  is a measure of the correlations between them \ct{osborne02} and is usually quantified 
in terms of quantities like concurrence and negativity \ct{hill97,peres96}.
For a transverse field Ising model concurrence has been found to maximize close to the QCP and  its derivatives 
show scaling behavior characteristics of that QCP \ct{osterloh02}. 
However, a different and  significant measure other than the entanglement, namely the  ``Quantum Discord" was introduced by Olliver and Zurek \ct{olliver01} which 
 exploits the fact that different quantum analogs of equivalent classical expressions can be obtained because of the fact that a measurement perturbs a quantum system.
 This property enables us to probe the ``quantumness" of a system. 
Quantum discord which ideally is  a subject of interest in quantum information theory \ct{werner89, bennett96,horodecki01,nielsen00,vedral07,wootters01} 
has  been studied  for spin systems and also close to  QCPs \ct{olliver01,luo08,dillen08,sarandy09,pal11,maziero11} and thereby establishes a 
natural connection between these two fields. 
Very recently an experimental study to measure quantum discord using an NMR set up has been reported \ct{auccaise11}.
  
In this paper, we study the quantum correlations present in the final state of a one dimensional transverse $XY$ model after quenching 
the system through an Ising critical point between two spins separated by a lattice spacing $n$ and quantify it in terms of quantum discord.  In the process, we also investigate 
the classical correlations and the mutual information bewteen two spins in the final state and study their behavior as a function of the rate of quenching.
 We compare our observations with the behaviour of two spin entanglement in the final quantum state
following a similar quench  as reported in a recent study \ct{sengupta09}. 
We also investigate  quantum discord  following  a quantum quench across a multicritical point (MCP) along a linear path.  We note that similar quenching studies have been carried
out to establish the universal scaling relation of the defect density namely the Kibble-Zurek scaling \ct{zurek05,polkovnikov05} generated following critical \ct{damski05,dziarmaga05,
levitov06,mukherjee07,sen08,polkovnikov_rmp} quenches. In reference \ct{sengupta09}, it was established that  concurrence does also follow the same Kibble-Zurek
scaling relation as the defect density; this prediction was found to hold good also for quenching through a MCP in a later study \ct{patra11}. We attempt to address the same question 
related to the scaling of discord; our studies indicate a similar scaling. 
  
  The organization of the rest of the paper is as follows. In Sec. II, we quantify  discord in terms of classical correlations and quantum mutual information. In Sec. III, we compute quantum discord of transverse $XY$ model after quenching the system through critical and multicritical points. This is followed by a discussion on our main results in Sec. IV. We present some concluding remarks in Sec. V. 
  
  \section{Quantum Discord}
  Let us consider a classical bipartite system comprising of two subsystems $A$ and $B$. The information associated with the system is quantified in terms of Shannon entropy $H(p)$ where $p$ is the probability distribution of the system. The classical mutual information is defined as
  \be 
  I(p)=H(p^A) + H(p^B) - H(p),
\label{total}
  \ee
 where $H(p^i), i=A, B$ stand for the entropy associated with the subsystem  $i$; this can alternatively be expressed as 
  \be
  J(p)=H(p^A)-H(p|p^B),
  \ee
  where $
  H(p|p^B)= H(p)-H(p^B) $
  is the conditional entropy. In the quantum context, the classical Shannon entropy functional  gets replaced by the quantum von-Neumann entropy expressed in terms of the 
density matrix `$\rho$'
 acting on the composite Hilbert space.The natural quantum extension of Eq.~(\ref{total}) is given by
 \ba 
 I(\rho)=s(\rho^A)+s(\rho^B)-s(\rho).
 \label{MI}
 \ea 
The conditional entropy based on local measurement however alters the system. The measurement are of von Neumann type having a set of one dimensional projectors $ \{\hat{ B_k }\}$ that sum up to identity. Following a local measurement only on the subsystem $B$, the final  
state $\rho_k$ of the composite system, which is the generalization of the classical conditional probability, is given by
 \be   
 \rho_k=\frac{1}{p_k}(\hat{I}\otimes \hat{B_k})\rho (\hat{I}\otimes \hat{B_k}),
 \ee
 with the probability $p_k={\rm tr}(\hat{I}\otimes \hat{B_k})\rho(\hat{I}\otimes \hat{B_k})$ where $\hat{I}$ is the identity operator for the subsystem $A$. Quantum conditional entropy can be defined as $s(\rho|\{\hat{B_k}\})=\sum_k p_k s(\rho_k)$,
 such that the measurement based quantum mutual information takes the form $J(\rho|\{\hat{B_k}\})=s(\rho^A)-s(\rho|\{ \hat{B_k}\}).$
 This expression maximized based on the local measurement gives the classical correlation \ct{henderson01}. Hence we have 
 \be
 C(\rho)=max _ {\{\hat{B_k}\}}  J(\rho| \{\hat{B_k}\}).
 \label{classical}
 \ee
 This line of arguments provides us with
 two quantum analogs of  the classical mutual information: the original quantum mutual information $I(\rho)$ (Eq.(\ref{MI})) and the measurement induced classical correlation (Eq.(\ref{classical})). 
As introduced by Olliver and Zurek \ct{olliver01}, the difference between these two, i.e.
 \be
 Q(\rho)= I(\rho)-C(\rho)
 \label{quantum}
 \ee 
is the quantum discord which measures the amount of quantumness in the state.  
It is noteworthy that  $I$  represents the total information (correlation) whereas 
$C$ is the information gained about $A$ as a result of a measurement on $B$. If $Q=0$, we conclude  that the measurement has extracted all 
the information about the correlation between $A$ and $B$, on the other hand, a non-zero $Q$ implies  that the information can not be extracted by local measurement and 
the subsystem $A$ gets disturbed in the process, a phenomena not usually expected in classical information theory .

\section{The model and Pairwise correlations}
\begin{figure}[ht]
\begin{center}
\includegraphics[width=3.9cm]{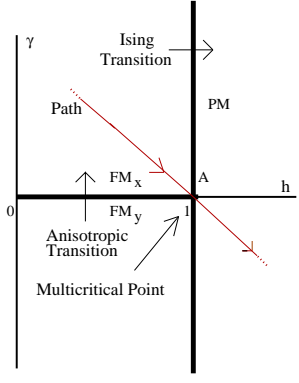}
\end{center}
\caption{(Color online) The phase diagram of one dimensional $XY$ model in a transverse field. The vertical bold line at $h= 1$ denotes Ising transition from ferromagnetic phase to paramagnetic phase. The horizontal bold line stands for anisotropic phase transition between ferromagnetic phase ordering in $x$ and $y$ directions. A linear path to approach the MCP `A' is shown.}
\label{Fig:phase}
\end{figure}

We  study pairwise correlations in a one-dimensional spin-1/2 $XY$ model in a transverse field with nearest neighbor ferromagnetic interactions described 
by the Hamiltonian \ct{lieb61,barouch70,bunder99,dutta10}
\be
H = - \frac{1}{2} ~\sum_i ~[(1+\gamma) \sigma^i_1 \sigma^{i+1}_1 + (1- \gamma) \sigma^i_2 \sigma^{i+1}_2 + h \sigma^i_3],
\label{h1} 
\ee 
where $\sigma$'s are the Pauli spin matrices and the subscript stand for the spin direction and superscript the lattice index. 
The parameter $h$ is the magnetic field applied in the transverse direction and $\gamma$ 
measures the anisotropy in the in-plane interactions; $\gamma=1$ refers to the transverse Ising model \ct{chakrabarti96}. 
The model can be exactly solved by mapping the spins to spinless fermions via a Jordan-Wigner transformation \ct{bunder99}; the phase diagram for the model is shown in Fig.~(\ref{Fig:phase}).

 We study the behavior of  quantum discord in the final state after quenching the system across an Ising critical point following the quench scheme $h(t)=t/ \tau$ with
$t$ going from $-\infty$ to $\infty$
\ct{levitov06}. The diverging relaxation time close to the QCPs at $h=\pm 1$ lead to defects in the final state. At $t \to -\infty$, the system is at the ground state $\ket 0$ where all the spins are aligned in the $+z$ direction.
 At $t\to \infty$ the system is in an excited state in which the probabilities of excitation for the mode $k$ is given by
 \be 
 p_k=\exp(-\pi \tau \gamma^2 \sin ^2 k).
 \label{pIsing}
 \ee
In the limit $\tau \to \infty$, only the modes close to the critical modes ($k=0$ or  $k=\pi$) contribute and one gets $p_k = \exp(-\pi \gamma^2 k^2 \tau)$.
 
 We further extend our study by quenching the system across a quantum multicritical point (MCP) by approaching along a linear path \ct{mukherjee10}
\begin{equation}
h(\gamma) = 1+|\gamma(t)| sgn(t);  ~~~\gamma(t)=-\frac{t}{\tau},
\label{path}
\end{equation} 
  and investigate the dependence of discord.
 The probability of defect formation $p_k$ \ct{divakaran09,deng09,mukherjee10}
  \be 
  p_k=\exp(- \pi \tau (1+\cos k)^{2} \sin ^2 k).
\label{pMCP}
\ee  

We shall now calculate various elements of  two-spin density matrix in the final paramagnetic phase of the Hamiltonian(\ref{h1}) for spins at the sites $i$ and $j=i+n$ using the
generic form of the  density matrix given by 
\ct{osborne02,syljuasen03,sarandy09}

\ba
 \rho^n = &\frac{1}{4}& (I^i \otimes I^j+c_1 \sigma_1^i \otimes \sigma_1^j + c_2 \sigma_2^i \otimes \sigma_2^j \nonumber \\
&+& c_3 \sigma_3^i \otimes \sigma_3^j 
 + c_4 I^i \otimes \sigma_3^j + c_5 \sigma_3^i \otimes I^j ),
\ea
 where 
   $c_1=\langle \sigma_1^i \sigma_1^{j} \rangle$, $c_2=\langle \sigma_2^i \sigma_2^{j} \rangle$, $c_3=\langle \sigma_3^i \sigma_3^{j} \rangle$,
 $c_4=c_5=\langle \sigma_3^i  \rangle$. For the Hamiltonian (\ref{h1}), the $X$ and $Y$ directions are equivalent and hence $c_1 = c_2$. The density matrix can also be expressed in the form
\begin{equation}
\rho^n = 
\left(
\begin{array}{cccc}
a^n_+ & 0 & 0 & b^n_1\\
0 & a^n_0 & b_2^n & 0\\  
0 & b^{n*}_2 & a_0^n & 0\\ 
b^{n*}_1 & 0 & 0 & a^n_-\\ 
\end{array}
\right),
\label{rdm}
\end{equation}
where the matrix elements are given in terms of the two-spin correlation functions in the following manner:
\begin{eqnarray}
a_{\pm}^n &=& \frac{1}{4} \langle (1 \pm \sigma^i_3)(1 \pm \sigma^{i+n}_3) \rangle = 1+c_3 \pm 2c_4, \nonumber \\
a_0^n &=& \frac{1}{4} \langle (1 \pm \sigma^i_3)(1 \mp \sigma^{i+n}_3)  \rangle =1-c_3, \nonumber \\
b_{1(2)}^n &=& \langle \sigma^i_- \sigma^{i+n}_{-(+)} \rangle. 
\label{elements}
\end{eqnarray}
We note that the up-down symmetry of the Hamiltonian simplifies the density matrix and some of the elements vanish \ct{syljuasen03}.
Defining a quantity  \ct{levitov06,sengupta09}
\begin{equation}
\beta_n = \int_0^{\pi} \frac{dk}{\pi} p_k \cos(nk),
\label{alpha_n}
\end{equation}
one gets
\begin{eqnarray}
c_4 = c_5 = \langle \sigma_3^i \rangle &=& 1 -2 \beta_0 ,\nonumber \\
c_3 = \langle \sigma_3^i \sigma_3^{i+n} \rangle &=& \langle \sigma_3^i \rangle^2 - 4 \beta_n^2.
\label{dcorr}
\end{eqnarray}
The expressions for $c_1$ and $c_2$ differ for different value of $n$; and it is presented below for $n \le 6$:
\ba
c_1& =& c_2 = 
\left\{
\begin{array}{lr}
\frac{\beta_2}{2}(1-2 \beta_0) &, n=2, \\
\\
(1-2 \beta_0)^2  \beta_2^2 - 4 \beta_2^4 + \non \\
\frac{\beta_4}{2} (1-2 \beta_0)^3 - 2 \beta_2^2 \beta_4  (1-2 \beta_0)&, n=4,\\
\\
\frac{1}{2}[ \beta_6 \{ (1 - 2 \beta_0)^2 - 4 \beta_2^2) \} + 4 \beta_2 \{ \beta_2^2 + \non \\
  \beta_4^2 - \beta_4 (1 - 2 \beta_0)\}] \times [16 \beta_2^2 \beta_4 +\non \\
 (1 - 2 \beta_0) \{ (1 - 2 \beta_0)^2 - 8 \beta_2^2 - 4 \beta_4^2\}] &, n=6. \nonumber \\
\end{array} 
\right.
\ea 
The eigen values of the density matrix are obtained in terms of the correlators $c_i$s
  \ct{sarandy09,luo08} as
\ba
  \lambda_0 &=& \frac{1}{4}[(1+c_3)+\sqrt{4c_4^2+(c_1-c_2)^2}], \nonumber \\
  \lambda_1 &=& \frac{1}{4}[(1+c_3)-\sqrt{4c_4^2+(c_1-c_2)^2}], \nonumber\\
  \lambda_2 &=& \frac{1}{4}[(1-c_3)+(c_1+c_2)], ~~\rm{and} \nonumber \\ 
  \lambda_3 &=& \frac{1}{4}[(1-c_3)-(c_1+c_2)],
  \label{eigenval}
  \ea
which can be expressed entirely in terms of $\beta$'s using the equations (\ref{elements}), (\ref{alpha_n}) and (\ref{dcorr}).

\section{Results}
In this section, we present results of the pairwise correlations in the final state of the spin chain as a function of the quenching rate $\tau^{-1}$. 
Using Eq.~(\ref{alpha_n}) we note that $\beta_n=0$  for odd $n$ as $p_k$ is invariant under $k \to \pi - k$. At the same time, 
$\langle \sigma_{\pm}^i \sigma_{\pm}^{i+n} \rangle= b_1^n=0$ for all $n$ since the expectation values  of a pair  of fermionic annihilation or creation operators do
always vanish. Moreover, $\langle \sigma_{\pm}^i \sigma_{\mp}^{i+n} \rangle= b_2^n=0$ for odd $n$ since the quantities $b_2^n$ are odd under the $\mathbb{Z}_2$ 
transformation\ct{sengupta09, levitov06}. On the other hand, $b_2^n= c_1+c_2$ for even $n$. 

The variation of mutual Information $I$, the classical correlation $C$ and the 
quantum discord $Q = I-C$ with $\tau$ are therefore studied for both critical and multicritical quenches (\ref{path}) for even $n$.
Let us rename the spin $i$ as subsystem $A$ and $j$ as subsystem $B$. The reduced density matrix for the subsystems $A$ and $B$  can be expressed as
 \ba
 \rho_A &=&  \frac{1}{2}( I^i \otimes I^j + c_4 I^i \otimes \sigma_3^j ), ~\rm{and} \nonumber \\
 \rho_B &=& \frac{1}{2}( I^i \otimes I^j + c_4  \sigma_3^i \otimes I^j  ).
 \ea
 with eigenvalues 
 \ba 
 \lambda_4 &=& \frac{1}{2}(1+c_4), ~\rm{and}\nonumber \\
 \lambda_5 &=& \frac{1}{2}(1-c_4).
 \ea
 The total mutual information  $I(\rho)$ is expressed terms of von Neumann entropies, which when substituted in Eq.~(\ref{total}) gives
 \ba 
 I(\rho)=s(\rho_A)+s(\rho_B)-\sum_{\alpha=0}^3 \lambda_\alpha \log_2 \lambda_\alpha , 
 \ea
 where
 $ 
 s(\rho_A)=s(\rho_B)=-\lambda_4 \log_2 \lambda_4 - \lambda_5 \log_2 \lambda_5 
 $. To calculate the classical correlation, we introduce a set of projector for local measurement on the subsystem $B$ given by 
   $B_k=V\Pi_kV^\dagger$ where $\Pi_k=\ket k \bra k : k=+,-$ is the set of projectors on the computational basis
 $\ket+=\frac{1}{\sqrt{2}}(\ket 0 + \ket 1), \ket-=\frac{1}{\sqrt{2}}(\ket 0 - \ket 1)$ and $V \in U(2)$ where  $V$ is parametrized
over a  Bloch sphere given by
\ba
\left(
\begin{array}{cc}
\cos \frac{\theta}{2}  &  \sin \frac{\theta}{2} e^{-i\phi}\\ 
\sin \frac{\theta}{2} e^{i\phi}  &  -\cos \frac{\theta}{2}\\ 
\end{array}
\right),
\ea
where the polar angle $\theta$ lies between  $0$ and  $\pi$ and the azimuthal angle $\phi$ is from  $0$ to $ 2\pi$.
Following a technique used in the reference \ct{sarandy09}, we can obtain the classical correlation by maximizing
\be
C(\rho)= s(\rho_A)-s(\rho_{+}),
\ee
where $\rho_{+}$ is the density matrix for the outcome $\ket k=\ket +$. Below we summarize the final results; e.g., 
 for $n=2$, we get

\ba
 c_1& =& c_2 = \frac{\beta_2}{2}(1-2 \beta_0), \nonumber\\
~ c_3 &=& (1- 2 \beta_0)^2 - 4 \beta_2^2, 
    ~c_4 = 1 - 2 \beta_0.
    \label{n=2}
\ea

\ba 
c_1 &=& c_2 = \frac{1}{2}[ \beta_6 \{ (1 - 2 \beta_0)^2 - 4 \beta_2^2) \} \non \\
&+& 4 \beta_2 \{ \beta_2^2 + \beta_4^2 - \beta_4 (1 - 2 \beta_0)\}]  \non \\
 &\times & [16 \beta_2^2 \beta_4 + (1 - 2 \beta_0) \{ (1 - 2 \beta_0)^2 - 8 \beta_2^2 - 4 \beta_4^2\}], \nonumber \\    
     c_3 &=& (1- 2 \beta_0)^2 - 4 \beta_6^2,  ~  c_4 = 1 - 2 \beta_0. 
    \label{n=6}
\ea 
    
The exact expressions for mutual information and classical correlation for $n=2$ is given below. 
\ba
I &=& -2(1- \beta_0) \log_2(1 - \beta_0) \non\\  
    && + ((1 - \beta_0)^2 - \beta_2^2)  \log_2((1- \beta_0)^2 - \beta_2^2) \non \\
&-& 2\beta_0 \log_2(\beta_0) + (\beta_0^2 - \beta_2^2) \log_2(\beta_0^2 - \beta_2^2) \non \\
    && + \frac{1}{4}\{ 4 \beta_0(1 - \beta_0)
      + 4\beta_2^2 + \beta_2(1 - 2\beta_0)\} \times \non \\
      &&~~~~~~\log_2 \left[ \frac{1}{4}\{4 \beta_0(1 - \beta_0) + 4\beta_2^2 + \beta_2(1 - 2\beta_0)\} \right] \non \\
    && + \frac{1}{4}\{4 \beta_0(1 - \beta_0) + 4\beta_2^2 - \beta_2(1 - 2\beta_0)\} \times \non \\
    &&~~~~~~ \log_2\left[ \frac{1}{4} \{ 4 \beta_0(1 - \beta_0) + 4\beta_2^2 - \beta_2(1 - 2\beta_0) \} \right], \non \\
\label{info2} 
\ea
and
\ba
 C &=& -(1- \beta_0) \log_2(1 - \beta_0) - \beta_0 \log_2(\beta_0) \non \\ 
    && + \frac{1}{2} \left( 1 - (1 - 2\beta_0) \sqrt{ 1+ \frac{\beta_2^2}{4}} \right) \times \non \\
    &~&\log_2 \left[ \frac{1}{2}\{ 1 - (1 - 2\beta_0) \sqrt{ 1+ \frac{\beta_2^2}{4}} \} \right]\non \\
 && + \frac{1}{2} \left( 1 + (1 - 2\beta_0) \sqrt{ 1+ \frac{\beta_2^2}{4}} \right) \times \non \\
    &~&\log_2 \left[ \frac{1}{2}\{1 + (1 - 2\beta_0) \sqrt{ 1+ \frac{\beta_2^2}{4}} \} \right].
\label{class2} 
\ea
Similarly one can obtain the expressions for $n=4$ and $n=6$ using the appropriate equations.
We note that  $I$ and $C$ and hence $Q=I-C$ depends entirely on  $\beta$'s which are in turn dependent on the quench rate $\tau^{-1}$ through the defect density $p_k$.
 In deriving the above expressions, we have used $\gamma=1$, however,
qualitatively the mathematical form of eigenvalues presented in Eq.~(\ref{eigenval}) remain unaltered though $\beta$'s are modified for $\gamma \neq 1$.

\begin{figure}[ht]
\begin{center}
\includegraphics[width=7.9cm]{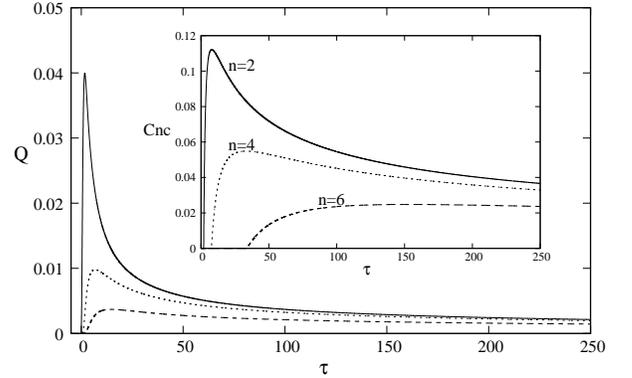}
\end{center}
\caption{ Quantum discord $Q$ as a function of $\tau$ $n=2$ (solid line),$4$ (dotted line) and $6$(dashed line) in the final state following a linear quench across the Ising critical point
with $\gamma=1$. Inset shows the variation of concurrence ($C_{\rm nc}$) for same parameter values as reported in \ct{sengupta09}.}
\label{Fig:QC}
\end{figure}

\begin{figure}[ht]
\begin{center}
\includegraphics[width=7.9cm]{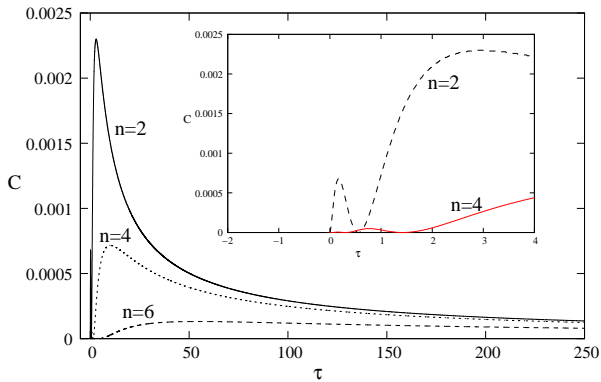}
\end{center}
\caption{Variation of classical correlation with $\tau$ for $n=2$ (solid line),$4$ (dotted line) and $6$(dashed line). Inset shows  small peaks for $\tau \to 0$ and it increases
 monotonically.}
\label{Fig:COsc}
\end{figure}

\begin{figure}[ht]
\begin{center}
\includegraphics[width=7.9cm]{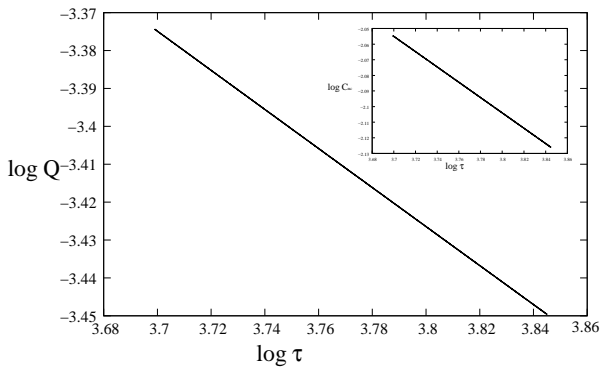}
\end{center}
\caption{ The variation of discord with $\tau$ for $n=2$ and $\gamma=1$ shown on a log-scale; the slope is $\approx -0.5$. Inset shows the variation of Concurrence with $\tau$
\ct{sengupta09} which shows a similar scaling.}
\label{Fig:isingcomp}
\end{figure}

 Fig.~(\ref{Fig:QC}) shows the variation of quantum discord $Q$ with $\tau$ for $n=2,4,6$  for a quench across the Ising critical point.
 As expected, $Q$ vanishes in both the limits $\tau \to 0$ and $\tau \to \infty$; the final state is nearly a direct product state in either
cases. discord initially increases with increasing $\tau$, and starts decreasing monotonically after reaching a peak at $\tau = \tau^m$. 
As $n$ increases, $\tau^m$ shifts towards the right. A similar behavior is observed for $I$. 
Fig.~(\ref{Fig:COsc})
shows that the classical correlations also exhibit a qualitatively identical variation with $\tau$ though it is smaller in magnitude in comparison to discord.
Surprisingly, the classical correlations however show some fluctuating behaviour for $\tau \to 0$  followed by the monotonic increase (see inset Fig.~(\ref{Fig:COsc})).
The value of $C$ is found to be one order of magnitude less than $I$ implying that correlations present in the system are mainly quantum mechanical.
 Though $C$ peaks at a larger $\tau$ in comaprison to  $I$ for a given $n$, it does not substantially influence  the behavior of $Q$ as quantum correlations 
apparently dominate  over classical correlations. 

\begin{figure}[ht]
\begin{center}
\includegraphics[width=7.9cm]{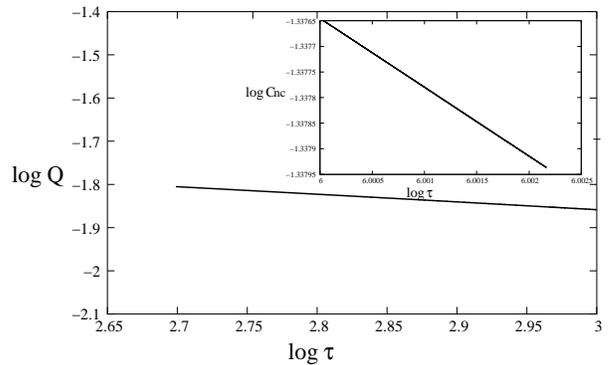}
\end{center}
\caption{The variation of discord for $n=2$ following a multicritical quench along a linear path with corresponding slope is
 $\approx -0.17$  which matches well with the value of the exponent $-1/6$ obtained for the defect density \ct{mukherjee10}. In the inset, we show the similar variation for concurrence with $\tau$ \ct{patra11} and the slope is $\approx -0.13$ which is in close agreement with the exponent $-1/6$.  
}
\label{Fig:multicomp}
\end{figure}

 The variation of concurrence in the final state for a quench across an Ising critical point with   $\tau$ has been studied recently \ct{sengupta09}
and the comparison is shown in Fig.~(\ref{Fig:QC}). Although the variation of discord and concurrence are qualitatively similar, we emphasize following differences.
The magnitude of discord is less than that of concurrence for the same $n$ by one order and it shows a peak at a value of  $\tau$ which is very small in comparison to the 
corresponding $\tau$ for concurrence. We conclude that the measurement based approach employed in calculating discord provides a quantitatively different result for
quantum correlations. Moreover, the study on concurrence \ct{sengupta09} indicates the existence of  a threshold value of $\tau$  above which the  bipartite entanglement is generated.
On the contrary,  discord is non-zero for all $\tau$ as we observe negligible shift close to $\tau=0$ for different $n$ as shown in Fig.~(\ref{Fig:QC}).  

We would now like to investigate the scaling of $Q$ or $I$ (since magnitude of $C$ is relatively smaller) as a function of $\tau$ like concurrence which was found to scale as 
   $1/{\sqrt \tau}$ in the limit of large $\tau$ for quenching through the Ising critical point \ct{sengupta09}. To explore the scaling of $I$ in the limit $\tau \to \infty$,
we analyse the asymptotic behavior of the terms present in Eq.~(\ref{info2}). While the first two terms taken together show a faster decay as $1/\tau$,  the total contribution
from the remaining four terms scales numerically as $1/\sqrt {\tau}$ and hence dominates when $\tau \to \infty$. Although, a closed power-law form is not obtained, our
studies apparently points to the fact that discord does also satisfy a scaling analogous to concurrence or defect density.

We further verify this claim by investigating the scaling of discord following a linear quenching across the MCP `A' (see Eq.~(9))
for which the defect density scales as $\tau^{-1/6}$  \ct{divakaran09,deng09,mukherjee10}.   In Fig.~(\ref{Fig:multicomp}), we compare the slope of discord with that of concurrence with respect to $\tau$ on a logarithmic scale and find a good 
matching which again indicates
that the scaling of discord is likely to be same as that of the defect density.

\section{Conclusion}
We have studied quantum discord and mutual information and their dependence on the quenching rate for two spins separated by $n$ lattice sites in the final state of a transverse $XY$ spin chain 
following a slow quench across a quantum critical and a multicritical point. 
Our studies show that the  quantum discord and  mutual information exhibit qualitatively similar behavior to that of concurrence as reported before \ct{sengupta09}; all these
quantities are in fact determined in terms of the eigenvalues of the  two-spin density matrix. However, we do also highlight differences; discord is
 smaller in magnitude in comparison to concurrence and 
is non-zero even in the limit of $\tau \to 0$ for any $n$ unlike concurrence which sets in at a threshold value of $\tau$ especially for large $n$.
The classical correlation is found to be smaller than the  mutual information 
 by one order of magnitude suggesting a stronger quantum nature of correlations. 
Finally, our studies
indicate that for a linear quenching through a MCP, discord shows a power-law scaling with the rate of quenching for large $\tau$ in the similar fashion as the defect density. This observation apparently suggests that the scaling of  discord in the final state in some sense is given in terms of that of the defect density for both critical and multicritical quenches though a closed power law form is not
obtained.

\begin{center}
{\bf Acknowledgements}
\end{center}
We acknowledge Victor Mukherjee and Diptiman Sen for comments. AD and AP acknowledge CSIR, New Delhi for financial support.


\begin{thebibliography}{99}

\bibitem{sachdev99} S. Sachdev, 
{\it Quantum Phase Transitions}(Cambridge University Press, Cambridge, England,1999). 

\bi{chakrabarti96} B. K. Chakrabarti, A. Dutta and P. Sen, {\it Quantum Ising Phases and transitions in transverse Ising Models}, m41 (Springer, Heidelberg,1996).

\bi{continentino} M. A. Continentino, {\it Quantum Scaling in Many-Body Systems} (World Scientific, 2001).

\bi{vojta03} M. Vojta, Quantum phase transition, Rep. Prog. Phys., vol. 66, p.2069. 

\bi{osterloh02} A. Osterloh, L. Amico, G. Falci and R. Fazio, Nature {\bf 416}, 608 (2002).

\bi{osborne02} T. J. Osborne and M. A. Nielsen, Phys. Revs. A, {\bf 66}, 032110 (2002).

\bi{amico08} L. Amico, R. Fazio, A. Osterloh, V. Vedral, Rev. Mod. Phys. {\bf 80}, 517-576 (2008).

\bi{guo05} S. J. Gu, G. S. Tian, H. Q. Lin, New Journal of Physics {\bf 71}, 052322 (2005).  

\bi{makhlin00} Y. Makhlin, G. Schoen and A. Shnirman, Rev. Mod. Phys. {\bf 73}, 357-400, (2001).

\bi{greiner02} M. Greiner, O. Mandel, T. W. Hänsch, and I. Bloch, Nature (London) {\bf 415}, 39 (2002).

\bi{jaksch98} D. Jaksch, C. Bruder, J. I. Cirac, C. W. Gardiner, and P. Zoller, Phys. Rev. Lett. {\bf 81}, 3108 (1998). 

\bi{hill97} S. Hill and W. K. Wootters, Phys. Rev. Letts. {\bf 78}, 5022 (1997); W. K. Wootters, {\it ibid} {\bf 80}, 2245 (1998); 
K. M. O'Connor, W. K. Wooters, Phys. Rev. A {\bf 63}, 052302 (2001).

\bi{peres96} A. Peres, Phys. Rev. Lett. {\bf 77}, 1413 (1996).

\bi{olliver01} H. Olliver and W. H. Zurek, Phys. Rev. Lett. {\bf 88}, 017901 (2001); W. H. Zurek, Rev. Mod. Phys. {\bf 75}, 715 (2003).
 
\bi{bennett96} C. H. Bennett, D. P. DiVincenzo, J. A. Smolin and W. K. Wootters, Phys. Rev. A {\bf 54}, 3824 (1996).

\bi{horodecki01} M. Horodecki, {\it Qunatum Inf. Comput.} {\bf I}, 3 (2001).

\bi{werner89} R. F. Werner, Phys. Rev. A {\bf 40}, 4277 (1989).

\bi{nielsen00} M. A. Nielsen and I. L. Chuang, {\it Quantum Computation and Quantum Information}(Cambridge University Press, Cambridge, UK, 2000).

\bi{vedral07} V. Vedral, {\it Introduction to Quantum Information Science}(Oxford University Press, Oxford, UK, 2007).

\bi{wootters01} W. K. Wootters, {\it Qunatum Inf. Comput.} {\bf I}, 27 (2001).

\bi{dillen08} R. Dillenschneider Phys. Rev. B {\bf 78}, 224413 (2008). 

\bi{luo08} S. Luo, Phys. Rev. A {\bf 77}, 042303 (2008). 

\bi{sarandy09} M. S. Sarandy, Phys. Rev. A, {\bf 80}, 022108 (2009).

\bi{pal11} A. K.  Pal and I. Bose, J. Phys. B: At. Mol. Opt. Phys. {\bf 44} 045101 (2011).

\bi{maziero11} J. Maziero, L. C. Céleri, R. M. Serra, M. S. Sarandy, arXiv:1012.5926 (Unpublished) (2011).

\bi{auccaise11}R. Auccaise, J. Maziero, L. C. Celeri, D. O. Soares-Pinto, E. R. deAzevedo, T. J. Bonagamba, R. S. Sarthour, I. S. Oliveira, R. M. SerraarXiv:1104.1596 (unpublished) (2011).

\bi{sengupta09} K. Sengupta and D. Sen, Phys. Rev. A {\bf 80}, 032304 (2009).

\bi{zurek05} W. H. Zurek, U. Dorner and P. Zoller, Phys. Rev. Lett.
{\bf 95}, 105701 (2005)

\bi{polkovnikov05} A. Polkovnikov, Phys. Rev. B {\bf 72}, 161201 (2005).

\bi{damski05} B. Damski, Phys. Rev. Lett. {\bf 95}, 035701 (2005).

 \bi{dziarmaga05} J. Dziarmaga, Phys. Rev. Lett. {\bf 95}, 245701
(2005); J. Dziarmaga, Advances in Physics {\bf 59}, 1063 (2010).

\bi{levitov06} R. W. Cherng and L. S. Levitov, Phys. Rev. A {\bf 73}, 043614 (2006).

 \bi{mukherjee07} V. Mukherjee, U. Divakaran, A. Dutta and D. Sen, Phys.
Rev. B {\bf 76}, 174303 (2007).

\bi{sen08} K. Sengupta, D. Sen and S. Mondal, Phys. Rev. Lett. {\bf 100}, 077204 (2008)  D. Sen, K. Sengupta and S. Mondal, Phys. Rev. Lett. {\bf 101}, 016806 (2008)

 \bi{polkovnikov_rmp} A. Polkovnikov, K. Sengupta, A. Silva and M. 
Vengalattore, arXiv:1007.5331 (2010).

 \bi{divakaran09} U. Divakaran, V. Mukherjee, A. Dutta and D. Sen, J.
Stat. Mech. (2009) P02007.

 \bi{deng09} S. Deng, G. Ortiz and L. Viola, Phys. Rev. B {\bf 80},
241109(R) (2009).

\bi{mukherjee10} V. Mukherjee and A. Dutta, EPL {\bf 92} (2010) 37004.

\bi{patra11} A. Patra, V. Mukherjee and A. Dutta (accepted for publication in J. Phys.: Conference Series) (2011).

\bi{henderson01} \bi{henderson01} L. Henderson and V. Vedral, J. Phys. A {\bf 34}, 6899 (2001); V. Vedral, Phys. Rev. Lett. {\bf 90}, 050401 (2003).

\bi{lieb61} E. Lieb, T. Schultz and D. Mattis, Ann. Phys. (NY) {\bf 16}, 407 (1961).

\bi{barouch70} E. Barouch, B. M. McCoy and M. Dresden, Phys. Rev. A {\bf 2}, 1075 (1970).

\bi{bunder99} J. E. Bunder and R. H. McKenzie, Phys. Rev. B {\bf 60}, 344 (1999).

\bi{dutta10}  A. Dutta, U. Divakaran, D. Sen, B. K. Chakrabarti, T. F. Rosenbaum and G. Aeppli, arXiv:1012.0653 (unpublished) (2010).

\bi{syljuasen03} O. F. Syljuasen, Phys. Rev. A {\bf 68}, 060301(R) (2003).  

\end{thebibliography}
\end{document}